\documentclass[11pt]{article}

\usepackage[english]{babel}

\usepackage{epsfig}
\usepackage{dcolumn}
\usepackage{amsmath}
\usepackage{changebar}
\usepackage{graphicx}

\newcommand{\ba} {\begin{eqnarray}}
\newcommand{\ea} {\end{eqnarray}}
\usepackage{graphicx}
\usepackage{dcolumn} 
\usepackage{bm}      
\usepackage{umlaut}
\begin{document}
\renewcommand{\figurename}{Fig.}
\renewcommand{\tablename}{Tab.}
\title{In-Medium Modifications of Hadron Properties
}

\author{ A.~Tawfik\thanks{tawfik@physik.uni-bielefeld.de} \\
 {\small Hiroshima University, 1-7-1 Kagami-yama, Higashi-Hiroshima 
 Japan }  
}

\date{}
\maketitle

\begin{abstract}
The in-medium modifications of hadron properties are briefly 
discussed. We restrict the discussion to the lattice QCD calculations for
the hadron masses, screening masses, decay constants and wave
functions. We review the progress made so far and describe how to
 broaden its horizon.  
\end{abstract}


\section{\label{sec:intr}Introduction}

''The origin of the mass'' is one of the fundamental questions in
physics. The mass of one composite system, like nucleus, is given by the
summation of the masses of its constituents plus the binding energy. The
proton and neutrons are composite systems too. But their masses can't be
obtained according to this principle. We have to assume that the mass is
mainly given by the kinetic energy of the quark and gluon fields. This
energy has to be smaller than the potential energy which confines the
qaurks in the hadron bound state. It is likely that the potential
energy can be modified in 
thermal and dense medium. This explanation for the origin of the mass
is likely valid in the chiral limit. As soon as the quarks get
bare masses, the chiral symmetry will be spontaneously
broken. Consequently, a mass difference appears between hadron states
with same spin but different parity quantum numbers. This part of the
mass will disappear, when the breaking of the chiral symmetry is
repealed at a critical temperature $T_c$. According to the lattice QCD,
the chiral symmetry breaking might partly be restored below
$T_c$. Therefore, the hadron masses are  conjectured to be modified at
finite temperatures and densities. 

Almost all hadron properties are extracted from
the exponential decay of the two-point correlation functions. In the
lattice QCD, the quark fields are discritized on the lattice sites. Any
resulting operator has to conserve the lattice symmetry. Such 
an operator can be projected into the different hadron
states. In present work, we restrict the discussion to the lattice QCD
calculations. We review the progress made so far. We discuss how to
extend this powerful tool to finite temperatures and densities. 

Understanding the modifications of the hadron properties at
finite temperatures and densities is very essential to study the
behavior of the ''hadronic'' matter under extreme conditions. It is also
important in order to interpret the phenomenology of hadron
production in heavy-ion collisions. It would not be possible to give a
solid description for the dynamics of the phase transitions, when 
these modifications are not taken into
account~\cite{TawNew}.    

The most extreme(st) modification occurs along the phase boundary. Not
only a rapid change in the degrees of freedoms is most likely expected,
but also a radical rearrangement of the correlation and entropy.

\section{\label{sec:mlatt}Hadron mass spectrum on the lattice}

\begin{figure}[thb]
\centerline{\includegraphics[width=8.cm]{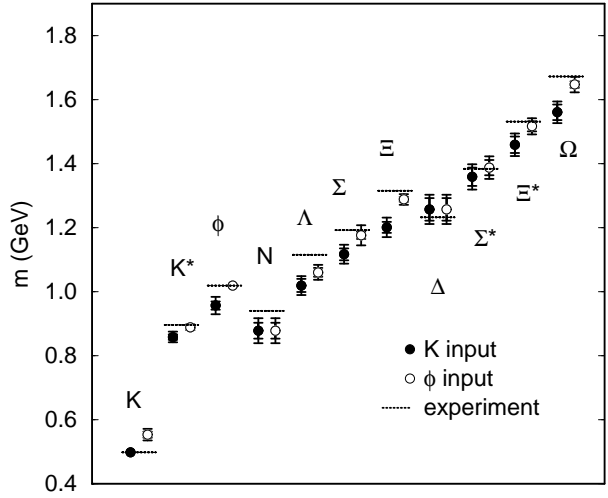}}
\caption{Comparison between lattice results on light
 hadron spectrum at zero temperature and vanishing density and the
 experimental values~\cite{cppacs02}. } \label{Fig:1}  
\end{figure}

We aim to investigate the ''origin of the mass''
from first principle. As mentioned above, all hadron properties
including the mass can be extracted from the two-point correlation
functions. The generic meson operators are given as:
\ba \label{eq:mstate}
{\cal O}_m(x) &=& \bar{\psi}^{a}_{\alpha}(x)\Gamma_{\alpha\beta}
\psi^{b}_{\beta}(x) 
\ea
where $\Gamma_{\alpha\beta}$ are the Dirac matrices
($1,\gamma_{\mu}, \gamma_5, \gamma_{\mu}\gamma_{5},
\sigma_{\mu\nu}$). The subscripts give the color quantum numbers. 
For the baryons 
\ba \label{eq:bstate}
{\cal O}_b(x) &=& \epsilon_{i j k } \left(\psi^{i,a}_{\alpha}(x)
                 \Gamma_{\alpha\beta} \psi^{j,b}_{\beta}(x) \right)
		 \psi_{\gamma}^{k,c} 
\ea
Then we can write down an expression for the two-point
correlation function and its exponential decay in the Euclidean
time. For pions, the operator needed to generate the spectrum reads 
\ba
P^{\dagger}(\vec{p},\tau) &=& \sum_{\vec{x}_n} e^{i\vec{p}.\vec{x}_n}
\bar{u}^a(\vec{x}_n,\tau)\gamma_5 d^a(\vec{x}_n,\tau) 
\ea
And at time $\tau$ the propagator is
\ba \label{eq:mpropg}
C(\vec{p},\tau) &\equiv& <P(\vec{p},\tau) P^{\dagger}(\vec{p},0) > =
\sum_h \frac{A_h(\vec{p})}{2 E_h(\vec{p})} \exp\left(-E_h(\vec{p})
\tau\right) 
\ea
where $A_h(\vec{p})$ represent of overlap of the operator from hadron
state $|h>$ with momentum $\vec{p}$. Below, we will show that the
decay constant is hidden in this parameter. Then the propagator can be 
integrated over the Euclidean time- and two spatial-directions.  
The mass will be extracted from $E(\vec{p})$. $E_h^2(\vec{p})=\vec{p}^2
+ m_h^2$ is the energy of hadron  state. An
example for  the hadron mass spectrum at zero momentum is given in
Fig.~\ref{Fig:1}. \\ 

\begin{figure}[thb]
\centerline{\includegraphics[width=6.cm]{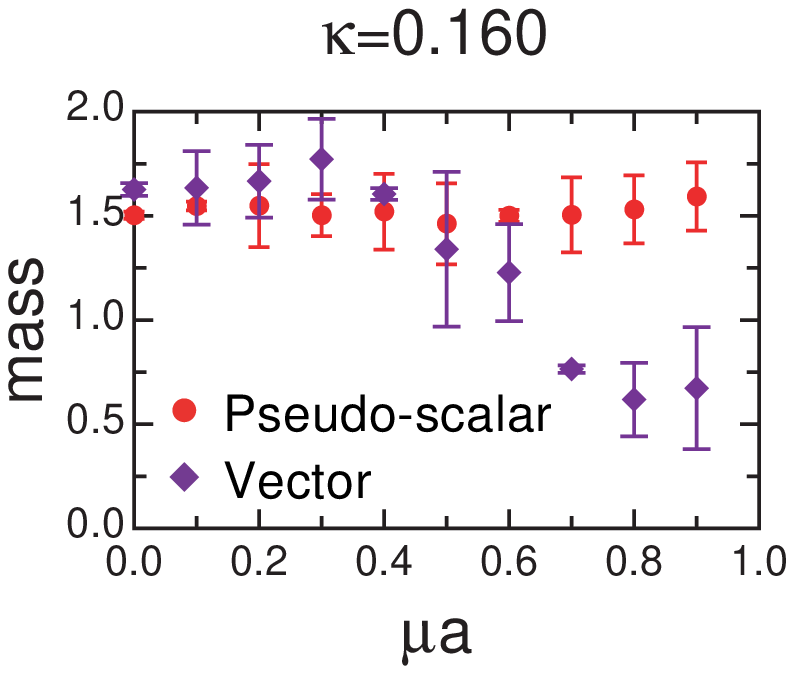}
\includegraphics[width=6.cm]{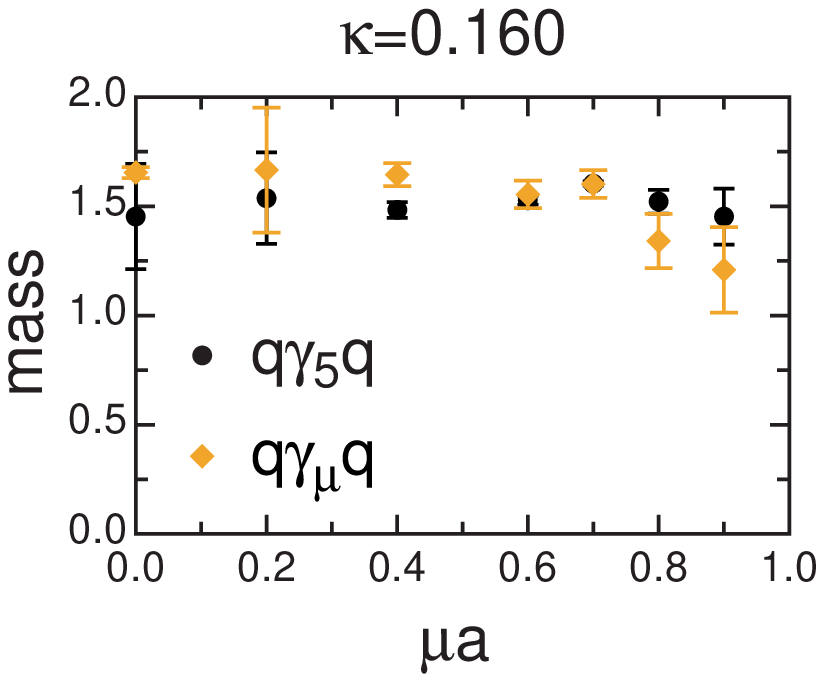}
}
\caption{Left panel: meson masses as function of chemical potential
 $\mu a$. The mass of pseudoscalar meson does not change with $\mu
 a$. The mass of vector meson drops with increasing $\mu a$. Right
 panel: the masses of baryon-type states are depicted in dependence
 on $\mu a$. No special behavior can be observed here.}
 \label{Fig:NakamMMass}  
\end{figure}

To extend these calculations over the region of finite temperatures, we
need temporal two-point correlation functions.
\ba \label{eq:temp1}
<{\cal O}_m(x)> &=& \frac{1}{{\cal Z}} \int d\psi d\bar{\psi} dU
\left(\bar{\psi}^a(x) \Gamma_{\alpha\beta} \psi^a(x)\right) \left(
\bar{\psi}^a(0) \Gamma_{\alpha\beta} \psi^a(0)\right)^{\dagger} \nonumber \\
 && \exp\left(- \bar{\psi} {\cal M} \psi -S_G\right) \\ \label{eq:temp2}
<{\cal O}_m(x,\tau)> &=& \frac{1}{{\cal Z}} \int dU\;
\Gamma_{\alpha\beta} {\cal 
 M}^{-1}(x,0) \;
 \Gamma_{\alpha\beta}^{\dagger} {\cal M}^{-1}(0,x) \;
 \exp\left(-S_G\right)  
\ea  
where ${\cal M}=\gamma_{\mu}D_{\mu}-m_q$ is the fermion determinant and
$S_G$ is the gauge action. $U$ stands for the link variable on the
lattice. 

The lattice calculations at finite chemical potential $\mu_q$ are not
trivial, since the fermion determinant becomes complex and therefore the
MC simulations are no longer applicable. Introducing imaginary chemical
potential can solve this problem. We can also apply the Taylor
series of the operator of interest at zero $\mu_q$.
\ba 
{\cal O}(\mu_q) &=& \left.C\right|_{\mu_q=0} + \mu_q \left.\frac{d
                    C}{d\mu_q}\right|_{\mu_q=0} \frac{\mu_q^2}{2}
		    \left.\frac{d^2 
                    C}{d\mu_q^2}\right|_{\mu_q=0} +\cdots 
\label{eq:mytaylor}
\ea

The first estimation for the hadron masses at finite chemical potential
$\mu$  is reported in Ref.~\cite{NakaMass}. The numerical study of
two-color QCD with  Wilson fermions on $4^3\times8$ lattice shows that
the mass of the pseudoscalar meson is almost constant at finite
$\mu$. The vector meson shows a different behavior. The mass of rho-meson
decreases with increasing $\mu$. This behavior
qualitatively agrees with the predictions from the effective
models, section~\ref{sec:hic}. In the baryonic channel, there is 
no change observed. It is important to note that these
results must be taken as a qualitative indication about the
behavior of the masses at finite chemical potential. As it was not
possible to define the physical units, both mass and $\mu$ are given in
units of lattice spacing $a$.

\section{\label{sec:smlatt}The Debye screening mass}

\begin{figure}[thb]
\centerline{\includegraphics[width=8.cm]{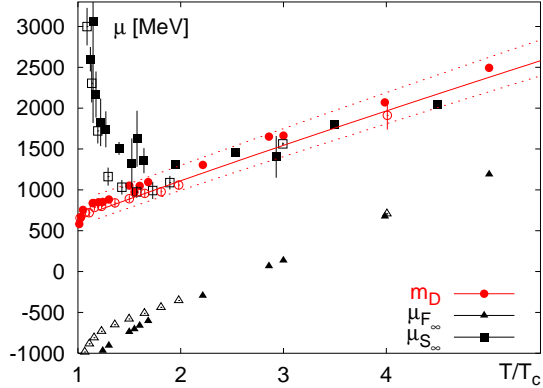}}
\caption{The Debye screening mass, $m_D(T)$ for 2-flavor (open
 circles) and quenched QCD (solid circles) as function of
 $T/T_c$~\cite{Zantw}. Different mass scales are also 
 included. } \label{Fig:2} 
\end{figure}

In contrast to the temporal correlation functions from which we construct
the hadron masses and decay constants, the spatial correlation functions
are often analyzed on the lattice. The reason is the limited  time
direction on lattice. The screening masses can be constructed from
the spatial correlation functions. At low temperatures, the screening
masses are identical to the pole masses. At high temperatures,
they are different. At very high temperatures, the screening masses 
are expected to approach $n\pi T$. $n\pi T$ gives the lowest Matsubara
frequency. 

The screening mass measures different quantities. It depends on the
properties of the many-body system. Therefore, it is a useful tool to
invistigate the plasma phase. The 
response of the medium to the existence of a weak perturbation, like
hadron, appears in form of  dynamical Debye screening of the
long-range Coulomb potential. There are different definitions for the
screening mass. The most common one relates it to the inverse Debye
screening length. Also it is defined as the static limit of the
longitudinal vacuum polarization tensor
$\prod_{00}(p\rightarrow0)=M_D$. 

The screening mass on lattice can be obtained from the exponential
falloff of a correlator of two Polyakov loops in the spatial
extension. The latter is related to the free energy in the limit
$x\rightarrow \infty$.  
\ba\label{eq:F}
\Delta F(x,T) &\propto&  \frac{\alpha_s(T)}{x}
                         \exp\left[-M_D(T)\;x\right] 
\ea
where $\alpha_s(T)=g^2(T)/4\pi$ is the strong coupling. Because
of the interaction with the medium via gluon exchange, the gluon
acquires an additional mass, called chromoelectric mass $mCE$.  $mCE$
results from the finite infrared limit of the polarization tensor. It
is proportional to $g(T)T$ in leading order perturbation theory.

As mentioned above, Eq.~\ref{eq:mpropg} can be extended to finite
temperature by adding temperature-depending terms. The integral of the
resulting correlation function over the Euclidean time- and two
spatial-dimensions results in the screening mass. 
\ba
E^2_h(\vec{p},T) &=& m_H^2 + \vec{p}^2 + \prod(\vec{p},T)
\ea 

Fig.~\ref{Fig:2} depicts the results on the screening mass
for quenched and 2-flavor QCD at zero chemical potential. The free
energy given in Eq.~\ref{eq:F}, can be separated into internal energy and
entropy. One finds that
the internal energy of this quark-antiquark system is related to the screening
mass. The dependence of screening mass on the temperature
$M_D(T)$ measures the strength of the screening response of the
medium due to inserting hadron in it. Different mass scales are also
included. We find that the effects of the hadron mass on the screening response
are much weaker than the effects of the energy and coupling. \\

In the perturbation theory, the screening mass - next-to-leading order -
reads 
\ba
m_D(T) &=& m_D^{(LO)} +  \frac{N g(T)T}{4\pi} 
          \ln \left(\frac{m_D^{(LO)}}{g^2(T)T}\right) + c_N g^2(T)T +
          {\cal O}(g^3(T)T) \\
          m_D^{(LO)} &=& \left(\frac{N_c}{3}+\frac{N_f}{6}\right)^{1/2} g(T)T
\nonumber 
\ea
where $c_N$ is a non-perturbative parameter. Therefore, it  can be
determined by non-perturbative methods, i.e. lattice.

Including the chemical potential is a non-trivial
task~\cite{Arnld}. In leading order approximation, the screening mass at
$\mu_q\neq0$ takes the form
\ba
m_D(T,\mu_q) &=& \left(N_c/3 + N_f/6+ (\mu_q/T)^2N_f/2\pi^2
\right)^{1/2}\;g(T)T 
\ea

\begin{figure}[thb]
\centerline{\includegraphics[width=10.cm]{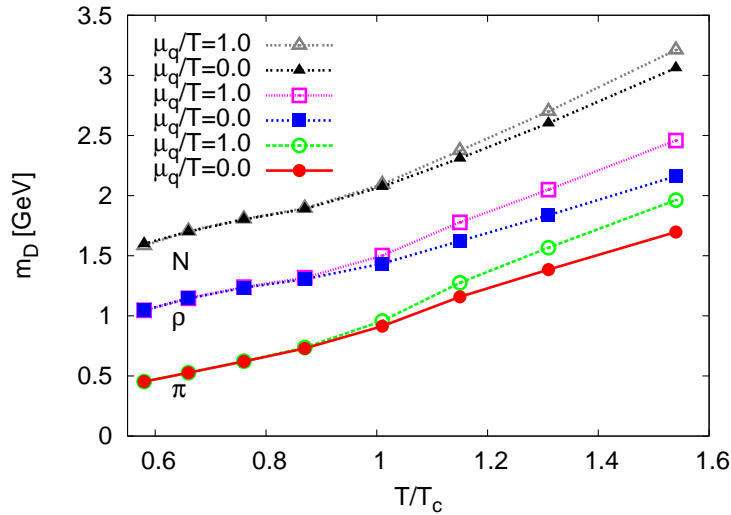}}
\caption{The screening mass, $m_D$  as function of $T/T_c$. Three
 hadron states are shown here. For each hadron we depict the results at
 $\mu_q/T=0.0$  and  $\mu_q/T=1.0$. We set the physical units according
 to $T_c=0.225\;$GeV. } \label{Fig:qcdtaro} 
\end{figure}

Alternatively, we can use the non-perturbative methods to calculate
$m_D(T,\mu_q)$. We can use the response of the screening mass at
$\mu_q=0$~\cite{qcdtaro1}. As in Eq.~\ref{eq:mytaylor}, we can 
apply Taylor expansion at fixed temperature, coupling and bare quark
mass to calculate $m_D(T,\mu_q)$
\ba
\frac{m_D(T,\mu_q)}{T} &=& \left.\frac{m_D(T)}{T}\right|_{\mu_q=0} +
          \left.\left(\frac{\mu_q}{T}\right)\frac{\partial
	  m_D(T)}{\partial\mu_q} 
          \right|_{\mu_q=0} +
	  \left.\left(\frac{\mu_q}{T}\right)^2 \frac{T}{2} 
           \frac{\partial^2 m_D(T)}{\partial\mu_q^2}\right|_{\mu_q=0} + \cdots
\label{Eq:Taylor}
\ea
The coefficients are determined by measuring the hadron
propagators and their derivatives at $\mu=0$ by using the standard MC
method.   
\begin{eqnarray}\nonumber
C_\pi(z) & = & C_1\,\left(e^{-\hat{m}_1\,\hat{z}} +
                          e^{-\hat{m}_1\,(N_z - \hat{z})}\right) 
          +  C_2\,\left(e^{-\hat{m}_2\,\hat{z}} + 
                          e^{-\hat{m}_2\,(N_z - \hat{z})}\right) \\
                           \nonumber \label{cor}
C_\rho(z) & = & C_1^\prime\,\left(e^{-\hat{m}_1^\prime\,\hat{z}} +  
                            e^{-\hat{m}_1^\prime\,(N_z -
			    \hat{z})}\right) 
           +  C_2^\prime\,(-1)^z\,\left(e^{-\hat{m}_2^\prime\,\hat{z}} +
                            e^{-\hat{m}_2^\prime\,(N_z -
			    \hat{z})}\right) \\ \nonumber 
C_N(z) & = &
C_1^{\prime\prime}\,\left(e^{-\hat{m}_1^{\prime\prime}\,\hat{z}} +  
              (-1)^z\,e^{-\hat{m}_1^{\prime\prime}\,(N_z -
	      \hat{z})}\right)  
        + 
       C_2^{\prime\prime}\,\left((-1)^z\,e^{-\hat{m}_2^{\prime\prime}\,
			    \hat{z}} +      
                              e^{-\hat{m}_2^{\prime\prime}\,(N_z -
			      \hat{z})}\right)
\end{eqnarray}
$N_z$ is the length of the lattice in the $z$-direction. The results for
degenerate quark masses $am_q=0.05$ are summarized in
Fig.~\ref{Fig:qcdtaro}. We note that the screening mass in all  
hadron channels increases with increasing the temperature. This means
that the response of the heat path to the existence of the hadrons is
enhanced as the temperature raises. Above $T_c$ this behavior seems to
be stronger. 

The influence of chemical potential starts to set on above
$T_c$. Below $T_c$, there is almost to influence. In the
deconfined phase the screening mass at finite $\mu_q$ is larger
than its value at $\mu_q=0$. The mesons are much sensitive than the baryons. It
is important to note that we get different screening
masses depending on the hadron state. This means that the heat bath
(gluonic system) is polarized around the static charge (the hadron) and
its parton density seems to be modified, accordingly.    

\section{\label{sec:dclatt}The Decay Constant}

In the continuum theory, the pion decay constant is defined as
\begin{eqnarray}
\sqrt{2} f_{\pi} m_{\pi} &=& \left<0|\bar{u} \gamma_4 \gamma_5
					d | \pi^+(\vec{p}=0)\right> 
\end{eqnarray}
We can use the temporal correlation functions given in
Eq.~\ref{eq:temp2} to calculate the decay constant for the hadron of
interest. At large separations and low temperatures the pole 
contributions dominate the correlation function.
\ba \label{eq:tfh}
<{\cal O}(\vec{p},\tau)> &=& \frac{{\cal C}(\vec{p})}{2E_h(\vec{p})}
\frac{\cosh(E_h(\vec{p})(\tau-\tau_0))}{\sinh(E_h(\vec{p}) \tau_0)}
\ea
where ${\cal C}(\vec{p})$ is the residue of the current when
$\vec{p}\rightarrow0$. Applying the GMOR relation, we can
construct the pseudoscalar correlation function, for instance, 
\ba\label{eq:fpi}
<{\cal O}(\tau)>_{\pi} &=& f_{\pi}^2 \frac{m_{\pi}^2}{8 m_q^2}
         \frac{\cosh(m_{\pi}(\tau-\tau_0))}{\sinh(m_{\pi} \tau_0)} 
\ea

There are extensive lattice calculations for the decay constants of
different hadrons at zero temperature~\cite{decayT0}. This quantity is
essential for the 
construction of CKM (Cabibbo-Kabayashi-Maskawa) unitary triangle
(quark mixing matrix). Its  shape is directly related
to the properties of the weak interactions and the phenomenon of
CP-violation. \\

\begin{figure}[thb]
\centerline{\includegraphics[width=7.cm]
            {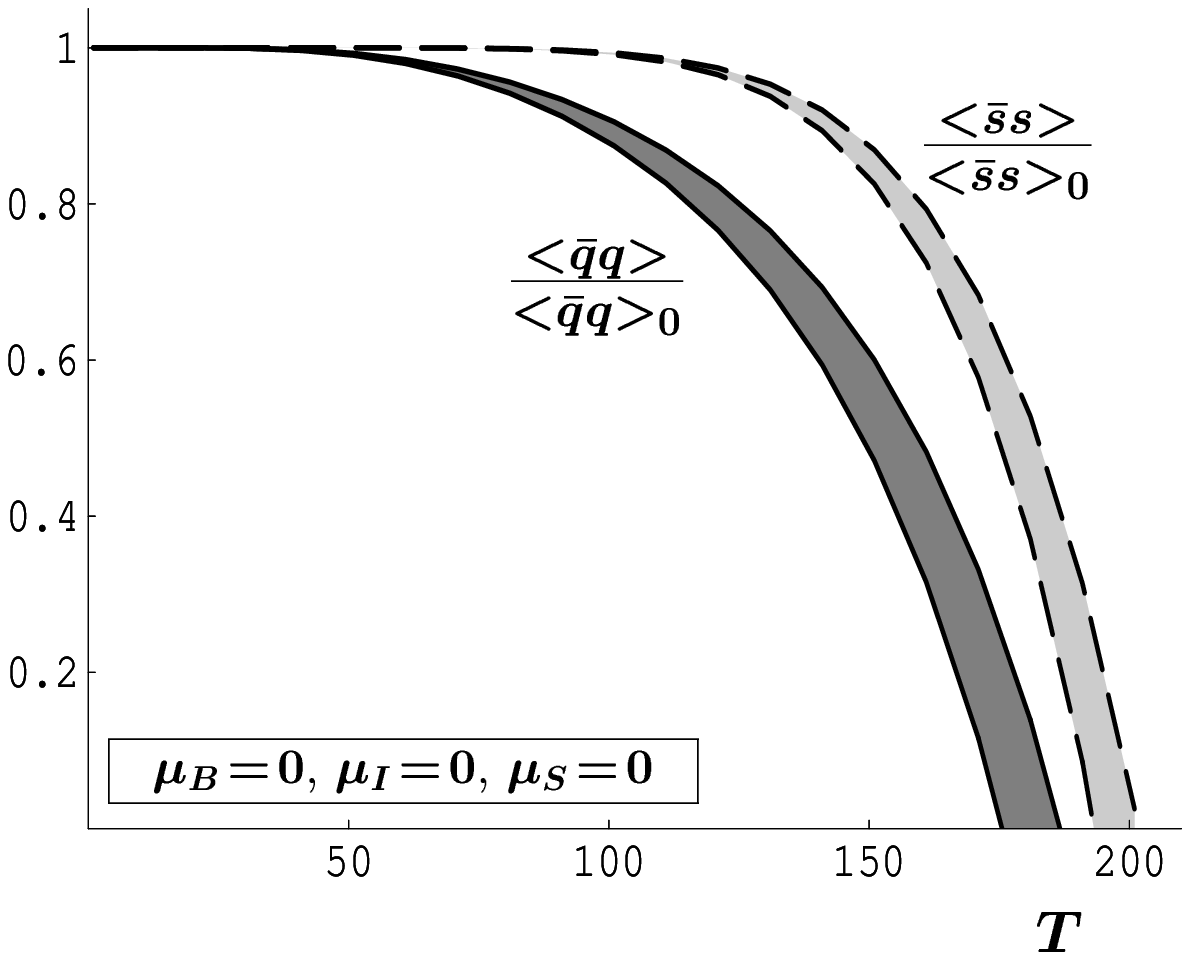}  
 \includegraphics[width=7.5cm]{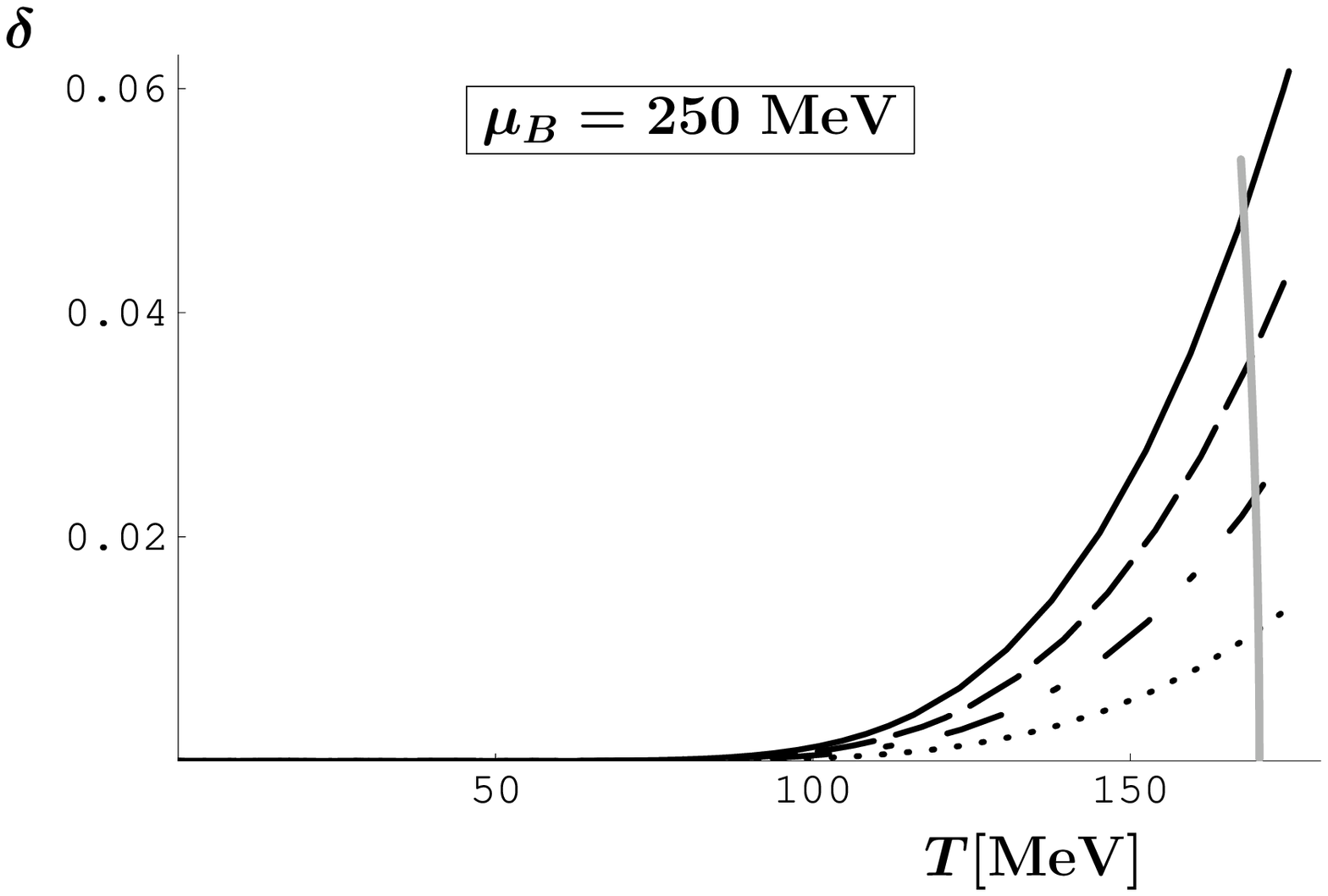}}
\caption{Left panel: light and strange quark condensate as function of
 temperature. Right panel: the difference between light quark
 condensates  normalized to the vacuum value calculated at
 a fixed baryochemical potential but different isospin chemical
 potentials. Depending on the medium, the quark condensates vanish 
 at different critical temperatures.} \label{Fig:qq} 
\end{figure}

Including chemical potential requires two types of modifications. The
first one deals with the influence of the finite density of the hadron
mass. This has been discussed above. The second one deals with the 
influence on the decay constant itself. 
\ba
(f_{\pi}^*)^2 (m_{\pi}^*)^2 = - 2 m_q <\bar{q}q>^*
\ea
The superscripts $*$ denote the in-medium modifications. $<\bar{q}q>^*$
is the quark condensates at finite temperature and density. In
Fig.~\ref{Fig:qq} we show the results from the resonance gas
model~\cite{qqCond}. At least according to the quark condensates, we can
easily expect that the decay constant has to be 
modified in the medium. To this end, we still need to study the in-medium
modification of the hadron masses. 

\section{\label{sec:dclatt}Wave function}

We assumed that the hadronic state given in
Eqs.~\ref{eq:mstate},~\ref{eq:bstate} have to fulfill the lattice
symmetry. But they still need a spatial dimension. The quarks have to be
positioned at a certian distance apart from each other. Even in the confined
phase, the quarks are not allowed to shear one location, since they are
fermions. Numerically,  
this requirement is convenient. We put the quarks at different lattice
sites separated by a distance $r$, for instance. The separated quarks
still have to remain connected to each other. To achieve this, we have
to add gluon links. The resulting signals for the ground state have to be
dominant in the contrast to that for the excited states~\cite{rajan}. For
example, the pion state can be expressed as
\ba
|\pi> &=& \alpha_1 |q\bar{q}> + \alpha_2 |q\bar{q}g> +  \alpha_3
|q\bar{q}gg> + \cdots
\ea
This is equivalent to adding smeared gauge fields. The added gauge
fields will build up a gluonic cloud around the quark-antiquark
system. What would this mean, technically? It means that one plaquette
has to be rescaled. And in the corresponding link variable a distance $r$
will appear, $\psi(\vec{x},r)$. Consequently, the operators,
Eqs.~\ref{eq:mstate},~\ref{eq:bstate}, have to be modified. From this
smearing procedure, the correlation functions becomes oscillating in 
$r$. The oscillation amplitude defines the wave function of the
corresponding hadron state. \\

The bound states can be studied by means of Bethe-Salpeter (BS)
function. The norm of the BS amplitude  \hbox{$|<0|{\cal 
O}(0)|h,\vec{p}>|$} is equivalent to the parameter $A_h^{1/2}(\vec{p})$
given in Eq.~\ref{eq:mpropg}. The BS amplitude gives the probability of
finding a pre-defined  state. This new (smeared) operator has to remain
gauge invariant. The gauge invariant  BS amplitude for
one pion with momentum $\vec{p}$ is

\ba
A_{\pi}(\vec{p},\vec{x}) &=& <0|\bar{d}(\vec{0}) \gamma_5 {\cal
G}(\vec{0},\vec{x}) u(\vec{x}) | \pi(\vec{p})>
\ea
where ${\cal G}(\vec{0},\vec{p})$ gives the path-ordered product of the
gauge links that connect the point $\vec{0}$ with $\vec{x}$.
The normalized two-point correlators construct the amplitude
\ba
A_{\pi}(\vec{p},\vec{x};\tau) &=& \frac{<0|\bar{d}(\vec{0};\tau)
\gamma_5 {\cal 
G}(\vec{0},\vec{x};\tau) u(\vec{x};\tau) \;\; \bar{u}(\vec{y};0) \gamma_5
d(\vec{y};0)| 0>}{<0|\bar{d}(\vec{0};\tau) \gamma_5 u(\vec{x};\tau)
\;\; \bar{u}(\vec{y};0) \gamma_5
d(\vec{y};0)| 0>}
\ea

From this procedure, it is now obvious,  that the wave function is also 
constructed from correlation function of two operators. The first
one represents the hadron state. The second one gives the annihilation
process. This picture is supported by the smearing process, as we
described above. Implementing this procedure under gauge
invariance assumption straightforwardly leads to calculating the wave 
function on lattice.  

The wave function is an essential quantity to investigate the bound
state. The influence of thermal and dense medium on the wave function
would reflect to what extend can the bound state survive. On the other
hand, one needs to elaborate the bound state potential in such a 
medium. We know from perturbative and non-perturbative calculations,
that the heavy quark potential gives a solid description for the
effective potential. It has been found that the effective potential of
two quarks with a short separation is a Coulomb-type. When the quarks are
pulled out to a very large distance, the effective potential becomes
linear.


\section{\label{sec:hic}Heavy-ion Collisions}

In this section, we discuss mass scaling laws which explain on how the
strong coupling change in dense and thermal medium.
Using effective Lagrangian which assumes spontaneously breaking of
chiral symmetry, Brown and Rho suggested a mass scaling in
medium~\cite{br91}, in order to explain the large mass enhancement of the
lepton pairs observed in CERES experiment.    
\ba
\frac{m_{\rho}^*}{m_{\rho}} &\approx& \frac{f_{\pi}^*}{f_{\pi}} \sim
         \left(\frac{<\bar{q}q>^*}{<\bar{q}q>}\right)^{1/2} 
\ea
Subsequently, the reduction of the masses of baryons and mesons are
equally scaled; $m_B^*=xm_B$ and $m_M^*=xm_M$. The validity of this
scaling law is restricted within the confined phase, especially in
the region after the freeze out. The gradual restoration of the chiral symmetry
breaking is essential to estimate this scaling law. 

Based on QCD sum rules Hatsuda and Lee~\cite{hl92}, suggested another
scaling law.  
\ba\label{eq:hl}
\frac{m_{\rho}^*}{m_{\rho}} &=& 1 - a \frac{n_B}{n_0} 
\ea
where $n_B$ is the baryon density and $n_0$ is the density of normal
nuclear matter. The parameter $a$ has a value ranging between $0.15$ and
$0.18$. 

From an effective theory with hidden local symmetry~\cite{harada}, it is
expected that the mass of vector meson entirely vanishes at $n_0$. At
this density, the pion decay constant will 
vanish too. Furthermore, a softening of the masses of $\sigma$ and $\rho$ 
mesons is expected to be associated with the gradual restoration of
chiral symmetry breaking.  \\

\begin{figure}[thb]
\centerline{\includegraphics[width=10.cm]{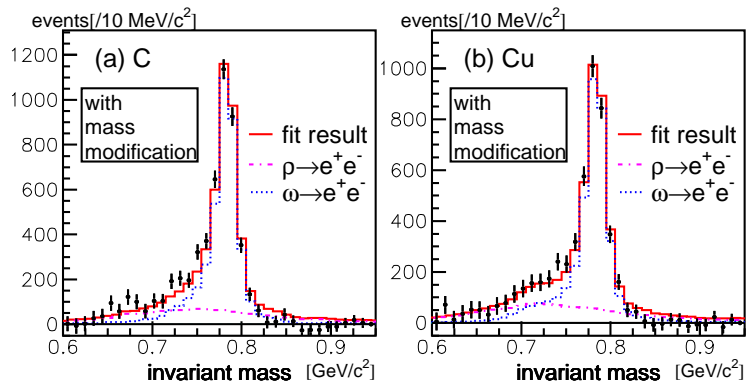}}
\caption{The invariant mass spectrum in two different targets (media) is
 measured at KEK. The observed access can only be explained, when
 the modification of masses in the medium is taken into
 consideration~\cite{invMRev} } \label{Fig:invM}  
\end{figure}

The results given in Fig.~\ref{Fig:invM} illustrate one application of
these models. The invariant mass and width modifications in heavy-ion
collisions are shown. These recent results from different nuclear media
obviously show an access. This access can only 
be explained, when the mass is scaled according to Eq.~\ref{eq:hl} with
$a=0.15$. The first direct measurement of the decay channel
\hbox{$\rho^0(770)\rightarrow\pi^+\pi^-$} (branching ratio $100\%$) is reported
in Ref.~\cite{Star}. Also a mass shift from $p+p$ to peripheral $A+A$ 
collisions has been observed. As a possible explanation for this
modifications one suggests the dynamical interactions with the
surrounding matter and the distortions of the correlations and phase space.

\begin{figure}[thb]
\centerline{\includegraphics[width=6.cm]{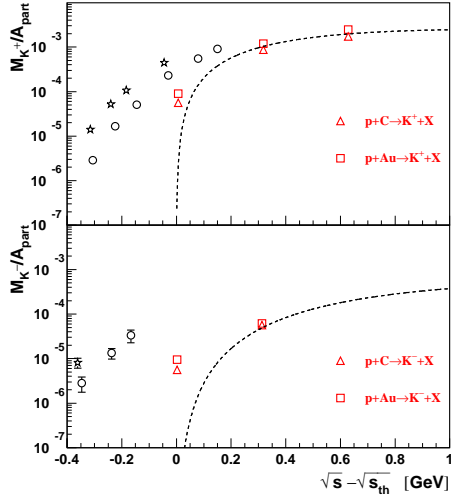}}
\caption{Particle production of $K^-$ (top panel) and $K^+$ (bottom
 panel) as function of the excess energy~\cite{kmpf}. The excess energy
 represents  the difference between $\sqrt{s_{NN}}$ and the energy required for
 $K^-$ ($K^+$) production. }
 \label{Fig:kmfr}  
\end{figure}

Another significant in-medium modification has been observed while
comparing the production and absorption in $p+p$, $p+A$ and
$A+A$ collisions. Fig.~\ref{Fig:kmfr} shows the strangeness production
and its in-medium modification. We observe an obvious  modification, when the
medium is changed. Going from $p+p$ to $A+A$ the medium density is
significantly increased. It is also interesting to note that the
modification depends on the hadron of interest. For $K^-$, the
modification is larger than for $K^+$. The reason is the different
influences of the medium on the particle production and absorption. 
The production of $K^-$ mainly depends on whether  there are hyperons in
the medium. The absorption of both kaons depends on the number of nucleons. 

\section{\label{sec:Summ}Summary}

We have discussed the in-medium modifications of the hadron
properties. We restricted the discussion to the lattice QCD calculations
for the hadron mass, Depye screening mass, decay constant and wave
function. We have given a brief review of the progress made so far and
discussed how to extend the lattice calculations to finite temperatures
and densities.

There are precise lattice calculations for the masses of low-lying
hadron 
states at zero temperature. Studying the thermal effects on the hadron
mass is essential to understand the hadron production and
absorption in heavy-ion collisions. Should the hadron mass fulfill
the description given by Hagedorn in 1960's and by the dual models in the
beginning of 1970's, our understanding of the phase diagram and the
transition at high temperatures and densities has to be revised (modified). The
dynamics controlling such a phase transition will be different, if the hadron
resonances shall keep or lose their finite volumes (finite wave
functions) and if their masses shall be modified or remain unchanged. We
might need to mention 
here that the hadron 
resonance gas calculations~\cite{Thrgm} did not assume finite volume and
mass scaling in dense and thermal medium.   

The screening mass gives the response of the medium to the existence of
a weak perturbation like the statically charged hadron. On the other
hand, this quantity describes the properties of the medium. We get
information on the parton density, entropy production and the long-range
correlations in the medium. As we have seen, there are different lattice
calculations using different approaches. We still need to study the
medium response to the low-lying parity partners, like $\sigma$ for
$\pi$, $a_1$ for $\rho$ and $N(1536)$ for $N$. Also we need
to study the screening mass of deconfined quarks in thermal and dense
plasma.   

We have also discussed the decay constant. To the auther's knowledge
there is no 
lattice calculations at finite temperature. At least there is no
systematic study for the thermal effects. Using the decay constant, we
can study the restoration of the chiral symmetry breaking. On the other
hand, the quark condensates in vacuum~\cite{TawQ} are obviously modified
in the medium. Including chemical potential is not trivial, because the
temporal correlation function is related to the
decay constant and simultaneously to the hadron mass. Both quantities
have to be expressed as function of chemical potential.

The wave function provides essential information on the bound
state. There is a framework to proceed calculations for the wave
function on lattice. We have mentioned that the wave function can be
expressed by the Bethe-Salpeter amplitude. Studying the hadronic bound
states at  finite temperatures and densities elaborates the possible
interactions with the medium and to what extend can the bound state survive. 

Finally we have discussed some phenomenological examples on the modification
of some hadron properties.

\vspace*{.5cm}

\noindent
{\bf Acknowledgment}\\ 

I would like to thank the Yukawa
Institute for Theoretical Physics at the Kyoto 
University where I have introduced this work at the YITP workshop on
``Hadrons ant Finite Density 2006'' 
YITP-W-05-24, Feb. 20-22, 2006. This work has been financially supported by 
the Japanese Society for the Promotion of Science. 

\end{document}